\documentclass[twocolumn,showpacs,amsmath,amssymb,prc,aps,10pt]{revtex4}

\usepackage{graphicx}
\usepackage{float}
\usepackage{dcolumn}
\usepackage{bm}
\usepackage{multirow}
\usepackage{amssymb}
\usepackage{amsmath}
\usepackage[usenames,dvipsnames]{color}

\newcommand{\GeV}{\ensuremath{\,{\rm GeV}}}

\newcommand{\rf}[1]{figure~\ref{#1}}

\begin{document}
\topmargin -1.0cm

\preprint{}

\title{Universal hadronization condition in heavy ion collisions at $\sqrt{s_\mathrm{NN}}= 62$ GeV and at $\sqrt{s_\mathrm{NN}}=2.76$ TeV}

\author{Michal Petr\' a\v n}
\author{Johann Rafelski}%
\affiliation{%
Department of Physics, The University of Arizona, Tucson, Arizona 85721, USA 
}%

\date{April 25, 2013}

\begin{abstract}
We obtain a detailed description of all available hadron multiplicity yields in central Pb--Pb collisions at {\small LHC} measured in the rapidity interval  $|y|<0.5$. We find that the  hadronization of the fireball at {\small LHC} occurs at nearly identical intensive physical bulk conditions for all centralities similar to those seen already at {\small RHIC}.
\end{abstract}

\pacs{25.75.Nq, 24.10.Pa, 25.75.-q, 12.38.Mh}
\maketitle


\noindent\textbf{Introduction and motivation:} We extend the successful description of central rapidity particle yields in a single freeze-out model~\cite{Rybczynski:2012ed,Broniowski:2001we}  to characterize the physical properties of the  hadronizing fireball. We consider, as an example, a supercooled quark--gluon plasma ({\small QGP}) disintegrating into hadrons which can scatter but preserve the stable particle abundance. Therefore hadron particle multiplicities  directly characterize the properties of the fireball. Final state hadrons are thus produced according to the accessible phase space with otherwise equal reaction strength. Accordingly, the particle yields are described by the chemical non-equilibrium statistical hadronization model ({\small SHM})~\cite{Torrieri:2004zz}. 

In this {\small SHM} implementation within the {\small SHARE}v2.2 program, the yields of particles are given the chemical freeze-out temperature $T$ and overall normalization $dV/dy$ (paralleling the experimental data  available as $dN/dy$). We include  phase space occupancies $\gamma_q,\gamma_s$ for light ($q=u,d$) and strange $s$ quark flavors, respectively, and we account for the small asymmetry between particles and anti-particles by fugacity factors  $\lambda_q,\,\lambda_s$ and the light quark asymmetry  $\lambda_{I3}$. These parameters enter the distribution function as $f_i(\varepsilon,T,\gamma_i,\lambda_i) = 1/(\gamma_i^{-1}\lambda_i^{\pm1} e^{\varepsilon/T} + S)$ for flavor $i$, where $S=-1,0,+1$ for bosons, Boltzmann distribution and fermions respectively. We refer the reader to Section 2 of~\cite{Rafelski:2007ti} for further discussion of the above parameters.

Our discussion addresses, evaluates and compares the {\small LHC} Pb--Pb experimental results available at $\sqrt{s_{NN}}=2760$\,GeV and {\small RHIC} Au--Au  at   $\sqrt{s_{NN}}=62.4$\,GeV.  We show that the chemical non-equilibrium {\small SHM} works at {\small LHC} in the 0--20\% centrality. This is so, since in the chemical non-equilibrium {\small SHM} approach, we allow quark pair yield parameter $\gamma_q>1$ for light quarks; this is the key difference from the simpler equilibrium  {\small SHM}.  The rationale for  $\gamma_q\ne 1$ originates in the high entropy density of  {\small QGP} at hadronization compared to the hadron phase space in which the color degree of freedom is frozen. In case that the fireball disintegrates faster than the time necessary to equilibrate the yield of light quarks bound in hadrons, the value $\gamma_q > 1$ must arise. 

In case that the fireball disintegrates faster than the time necessary to equilibrate the yield of light quarks bound in hadrons, either the value $\gamma_q > 1$  arises,  or  one must consider a dynamical volume growth as a path for absorbing the excess entropy of {\small QGP} source. However, this second option requires a much longer lifespan of the particle source than is supported by {\small HBT} data~\cite{Csernai:2002sj,Aamodt:2011mr} and thus is experimentally excluded: the observed LHC total lifespan ($\tau_f\simeq10\,\mathrm{fm}/c$~\cite{Aamodt:2011mr}) favors very fast, or sudden, hadronization~\cite{Csernai:1995zn,Rafelski:2000by}. In this situation, chemical non-equilibrium approach must be applied also to the light quark abundance, introducing the light quark phase space occupancy $\gamma_q$. This proposal made for the high energy {\small SPS} data~\cite{Letessier:1998sz,Letessier:2000ay} helped also to improve the understanding of {\small RHIC}200 hadron rapidity yield results~\cite{Rafelski:2004dp} and allowed a consistent interpretation of such data across the full energy range at {\small SPS} and {\small RHIC}200~\cite{Letessier:2005qe}.

{\small SHM} fits, that arbitrarily set  $\gamma_q=1$~\cite{Abelev:2012vx,Abelev:2012wca}, (equilibrium {\small SHM}), describe    hadron yields at  {\small LHC} with relatively large total $\chi^2$. This chemical equilibrium {\small SHM}  disagrees at {\small LHC} across many particle yields, but the greatest issue is the `proton anomaly', which makes it impossible to fit  the p$/\pi = 0.046\pm0.003$ ratio~\cite{Abelev:2012wca} along  with the multi-strange baryons $\Xi$ and $\Omega$.

We will show, that common intensive {\small QGP} bulk properties arise, nearly exactly equal to those found at {\small RHIC}, for all four collision centralities we analyze. We will discuss in depth the main extensive bulk property difference we find, that is the entropy  $dS/dy$ growth with energy and centrality.

\vskip 0.2cm
\noindent{\textbf{Fit to most central collisions:}} 
Within the chemical non-equilibrium   {\small SHM} we have, allowing for baryon-antibaryon asymmetry, seven independent statistical model parameters reduced by two constraints: a) The isospin fugacity factor $\lambda_{I3}$ is constrained by imposing the charge per baryon ratio, $(\langle Q \rangle \!-\! \langle\overline{Q}\rangle)/(\langle B\rangle \!-\! \langle\overline{B}\rangle)\simeq 0.38$, present in the initial nuclear matter state at initial instant of the collision; b) For each value of  $\lambda_q$, strangeness fugacity $\lambda_s$ is evaluated by imposing the strangeness conservation requirement $\langle s\rangle\!-\!\langle \bar s\rangle\simeq 0$.  Considering the particle--anti-particle symmetry at {\small LHC}, the four key  parameters are the hadronization volume $dV/dy$, temperature $T$ and the two phase space occupancies $\gamma_q $ and $\gamma_s$. The 7th parameter is  light quark fugacity $ \lambda_q$.

We use as input to our fit the  hadron yield data in 0--20\% centrality bin as presented in Ref.~\cite{QM2012b}, where a fit to this data set for the case of chemical equilibrium ($\gamma_s=\gamma_q=1$) is shown. For comparison and demonstration of method compatibility,  the chemical equilibrium model fit (dashed lines in \rf{fig:firstfit}) is shown, with a large   $\chi^2=34.8$. In both approaches, we fit the same data,  the decrease in  $\chi^2$ by factor nearly 5 is due to  chemical non-equilibrium i.e., $\gamma_q\neq1,\,\gamma_s\neq1$. We determine the best light quark fugacity factor $\lambda_q=1.00359$, which corresponds to baryo-chemical potential $\mu_B\simeq 1.5\,\mathrm{MeV}$, and we apply strangeness and charge per baryon conservation by fitting them as two additional data points. The result of our 0-20\% centrality bin fit is shown in \rf{fig:firstfit} and in top section of the third column of table~\ref{tab:parameters}.  We compare our present results to our recent analysis~\cite{Petran:2011aa} of Au--Au collisions at $\sqrt{s_{NN}}=62.4\GeV$ at {\small RHIC}62,  shown in the 2nd column in table~\ref{tab:parameters}.

\begin{figure}[!tb]
\includegraphics[width=0.9\columnwidth]{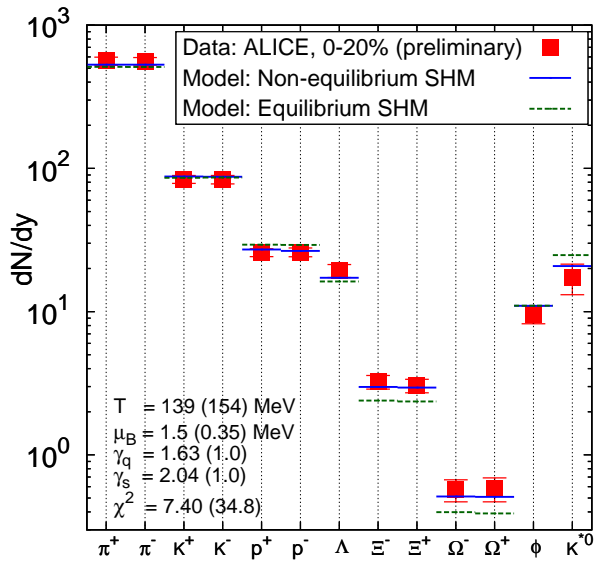}
\caption{\label{fig:firstfit}(color online) The non-equilibrium {\small SHM} fit is indicated by (blue) solid horizontal lines overlaying for all the  {\small  LHC-ALICE} (preliminary) data available in 0--20\% centrality bin (red squares). Chemical equilibrium fit is indicated by (green) dashed lines with model parameters presented in parentheses.}
\end{figure}

\begin{table*}[!tb]
\caption{\label{tab:parameters} Top section shows chemical non-equilibrium {\small SHM} fit parameters $dV/dy$, $T$, $\gamma_q$, $\gamma_s$ and $\chi^2_{\rm total}$ with ndf (number data less number of parameters) obtained in each centrality bin. Errors are a fit-stability estimate obtained with K$^\pm$ yield shifted within experimental error, the underlying statistical fit error is negligible. Bottom section presents fireball bulk properties in each bin: energy density $\varepsilon$, pressure $P$, entropy density $\sigma$, strangeness per entropy content $s/S$, and entropy at {\small LHC}2760 compared to {\small RHIC}62, $S_{\mathrm{LHC}}/S_{\mathrm{RHIC}}$. The centrality defining number of participants $N_{part}$ values are adopted from~\cite{Abelev:2013qoq}.}
\begin{ruledtabular}
\begin{tabular}{l||c||cccc}
	& \textbf{{\small RHIC}62\ \ } & \multicolumn{4}{c}{\textbf{{\small LHC}2760}}  \\ \hline
Centrality  & 0--5\% & 0--20\% & 20--40\% & 40--60\% & 60--80\% \\
$\langle N_\mathrm{part}\rangle$& 346 & 308 & 157 & 68.8 & 22.6 \\ 
\hline\hline & & & &\\[-0.28cm]
$dV/dy\,\mathrm{[fm^3]}$ & $853$  	& $2455\pm146$	& $1169\pm9$		& $406\pm3$		& $102\pm7$ 		\\[0.07cm]
$T\,\mathrm{[MeV]}$      & $139.5$	& $138.6\pm1.1$	& $137.6\pm0.03$	& $140.5\pm0.04$	& $143.2\pm0.08$		\\[0.07cm]
$\gamma_q$               & $1.58$ 	& $1.627\pm0.007$	& $1.633\pm 0.0002$	& $1.616\pm0.003$	& $1.60\pm0.02$		\\[0.07cm]
$\gamma_s$               & $2.24$ 	& $2.04\pm0.04$	& $2.01\pm0.12$	& $1.83\pm0.08$	& $1.70\pm0.09$		\\[0.07cm] 
$\chi^2_{\rm total}/{\rm ndf}$  & $0.38/5$ 	& $7.40/8$ 	& $2.93/5$	& $3.58/5$	& $5.43/5$	\\[0.07cm] 
\hline\hline & & & &\\[-0.28cm]
$\varepsilon\,\mathrm{[GeV/fm^3]}$	& $0.493$ & $0.466\pm0.018$ & $0.441\pm0.012$ & $0.488\pm0.010$ & $0.536\pm0.025$ \\[0.04cm]
$P\, \mathrm{[MeV/fm^3]}$        	& $82.0 $ & $79.1\pm2.8 $ & $75.5\pm1.5 $ & $82.2\pm1.3 $ & $90.2\pm4.0 $ \\[0.04cm]
$\sigma\, \mathrm{[fm^{-3}]}$     	& $3.40 $ & $3.23\pm0.11$ & $3.07\pm0.07 $ & $3.36\pm0.06 $ & $3.65 \pm 0.13 $ \\[0.04cm]
$s/S$                    			& $0.0322$ & $0.0296\pm0.0002$ & $0.0289\pm0.0014$ & $0.0277\pm0.0009$ & $0.0267\pm0.0011$  \\[0.04cm]
$S_{\mathrm{LHC}}/S_{\mathrm{RHIC}}$     & --- & $3.05$ & $2.66$ & $2.18$ & $1.52$  \\
\end{tabular}
\end{ruledtabular}
\end{table*}

\begin{figure}[!tb]
\includegraphics[width=0.98\columnwidth]{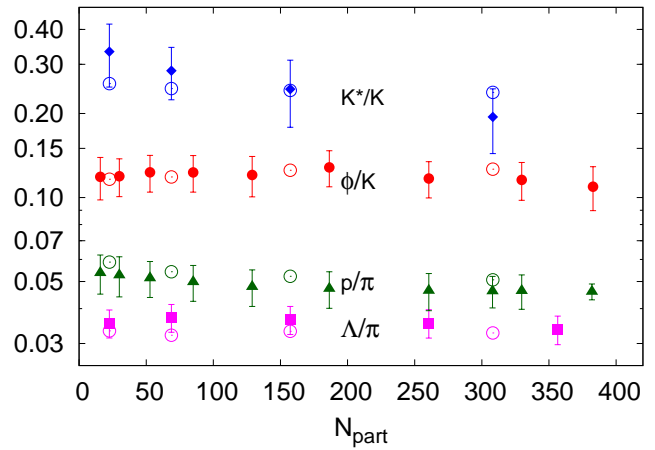}
\caption{\label{fig:ratios}(color online) Experimental data (full symbols) and model predicted (open circles) for particle ratios as a function of centrality. See text for discussion of data and results.}
\end{figure}

\vskip 0.2cm
\noindent\textbf{More peripheral centralities:} We extend our study to more peripheral collisions at {\small LHC} using much smaller data set, complemented by two assumptions as follows:\\
{\bf a)} We consider the three ratios K*$^0$/K$^-$, $\Lambda/\pi\equiv 2\Lambda/(\pi^-+\pi^+)$ and $\phi/$K$^-$, as presented in~\cite{Singha:2012qv} and which we show in \rf{fig:ratios} with full symbols. We fit these ratios in three centrality bins, 20--40\%, 40-60\% and 60-80\%, in which K*/K and $\Lambda/\pi$ have experimental data point. We take average of two neighboring $\phi/$K data points in order to use this ratio as input to our fit in the intermediate centrality. This is consistent with the claim that $\phi/$K is constant over all centralities and has been claimed independent of centrality in~\cite{Singha:2012qv}.\\
{\bf b)} To obtain overall normalization, we complement the ratios with charged particle rapidity density $dN_{\mathrm{ch}}/dy$. Based on our fit of 0--20\% centrality data, we see that the ratio of charged particle rapidity to pseudo-rapidity density is $(dN_{\rm ch}/dy)/(dN_{\rm ch}/d\eta)=1.115$. We multiply data from~\cite{Aamodt:2010cz} by this factor, and use the resulting $dN_{\rm ch}/dy$ as an additional data point, that determines the value of fireball  volume $dV/dy$. The input multiplicity data is presented in the \rf{fig:particle-predictions}.\\
{\bf c)} We include the yields of $\pi^\pm$, K$^\pm$ and p$^\pm$, as presented in~\cite{Abelev:2013vea} and shown in \rf{fig:particle-predictions}. Similarly to $\phi/$K ratio, we use as input averages of yields in two neighboring centrality bins. 

We fit the three centrality bins of {\small LHC}2760 data using the eight data points and the two conservation laws, which fix $\lambda_s$ and $\lambda_{I3}$. We thus have five degrees of freedom as is seen in table~\ref{tab:parameters} in 4th, 5th and 6th column. All studied centralities show reasonable  $\chi^2/\mathrm{ndf}\leq 1.1$. We find for all four  {\small LHC} centrality bins remarkably similar statistical parameters. 

We compare  the outcome of the fit showing  the three input ratios K*$^0$/K$^-$, $\Lambda/\pi$ and $\phi/$K$^-$, and p$/\pi$ in \rf{fig:ratios}. All data are well fitted including p$/\pi$, which we evaluate from the individual fitted yields of p and $\pi$ from~\cite{Abelev:2013vea}.

The predicted particle yields normalized by $N_{\rm part}/2$ are shown in \rf{fig:particle-predictions}, where considering particle--anti-particle symmetry only one of the isospin multiplets is shown. The $\pi,$K and p yields are fitted values, whereas the other particle yields are predictions. We indicate in \rf{fig:particle-predictions} on the right edge the experimental input for the  0--20\%  bin with an offset to assure visibility of the small differences between fit and experimental data. We show  the  input multiplicity $(dN_{\rm ch}/dy)/(N_{\rm part}/2) $ and fit result in \rf{fig:particle-predictions}. Both are overlapping exactly for the three peripheral bins since this is the most precise input data.

\begin{figure}[!tb]
\includegraphics[width=0.98\columnwidth]{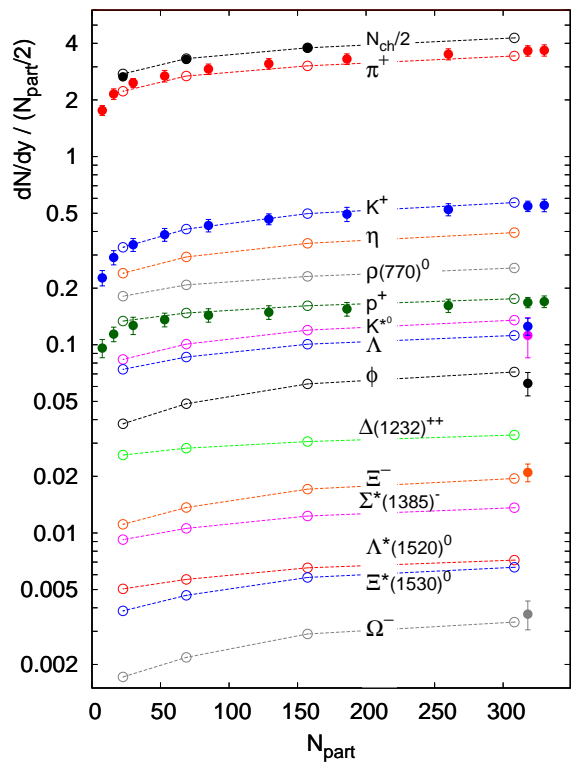}
\caption{\label{fig:particle-predictions}(color online) Predicted particle yields per participant pair as a function of centrality. Open symbols represent our model predictions; lines guide the eye. We  show experimental data for the 0--20\% bin as full symbols, with an offset.}
\end{figure}

\vskip 0.2cm
\noindent\textbf{Hadronization conditions:} Despite a change by a factor of 45 in reaction energy comparing {\small RHIC} and {\small LHC}, and the wide range of centrality,  the only quantity among statistical parameters shown in table~\ref{tab:parameters} that significantly changes is $dV/dy$. This suggests, that we should look closer at the intensive bulk physical properties of the fireball: The emitted hadrons not only carry away from the fireball the above  discussed   charge, baryon number or strangeness, but also e.g. the thermal energy,  $dE/dy$  obtained by summing the energy content of all produced particles, observed and predicted.  

The bulk thermal energy density at hadronization $\varepsilon$ defined by $\varepsilon \equiv (dE/dy)/(dV/dy)$ is of direct interest. Similarly, we evaluate the entropy  $dS/dy$,  pressure $P$ and total yield of strangeness $ds/dy\equiv d(q_s+\bar q_s)/2dy$.  These properties of the fireball at hadronization are shown in the bottom section of the table~\ref{tab:parameters}, where for simplicity we omit in the first column the symbol $d/dy$.  

As one can see by comparing  the second and third columns of table~\ref{tab:parameters}, the intensive properties of the  {\small RHIC}62 fireball at hadronization  i.e.  $\varepsilon,P,\sigma$ and $s/S$, are practically identical to the here evaluated case {\small LHC}2760, and this continues across all considered centralities as is seen in the 4th, 5th and 6th column of table~\ref{tab:parameters}. We show a comparison of $\varepsilon,P$ as a function of centrality between {\small LHC} (solid symbols) and {\small RHIC} (open symbols) in the bottom part of \rf{fig:entropy}. The difference between {\small LHC} and {\small RHIC} can be easily attributed to the fit uncertainties, since the intensive quantities are proportional to a high power of statistical parameters. 

Hadronization volume $dV/dy$ does not characterize the early stage of a collision, this information is available in the entropy content at hadronization $dS/dy\propto dV/dy$ which presents a more accurate view of the pre-hadronization processes that created the fireball. For an ideally flowing and expanding {\small QGP}, most of the observed entropy yield $dS/dy$ of a fireball is created early in the collision.  In the top panel of \rf{fig:entropy} we see that at {\small LHC}, $dS/dy \propto N_{part}^{\sim1.173}$, the rise is faster than linear. For comparison note that at {\small RHIC}, the entropy yield rises almost linearly with $N_{part}$. Last row of table~\ref{tab:parameters} shows the enhancement of entropy at {\small LHC} compared to {\small RHIC}, $S_{\mathrm{LHC}}/S_{\mathrm{RHIC}}$. The enhancement decreases as a function of centrality from $\sim 3$ to $\sim 1.5$, which implies additional entropy production mechanism proportional to centrality at {\small LHC}.

\begin{figure}[!tb]
\includegraphics[width=0.95\columnwidth]{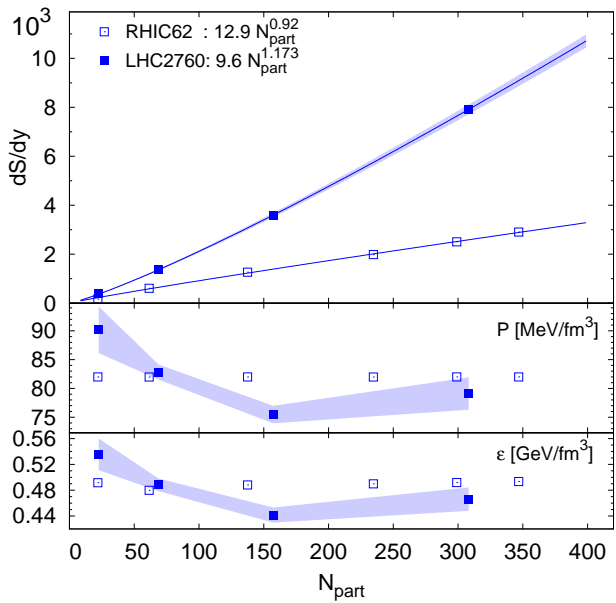}
\caption{\label{fig:entropy}(color online) Top panel shows entropy content of the fireball at LHC (full symbols) and RHIC (open symbols) as a function of centrality. Bottom part shows pressure $P$ and energy density $\varepsilon$ at hadronization with the same symbols for LHC and RHIC as in the top panel, for values see table \ref{tab:parameters}.}
\end{figure}

Strangeness per entropy $s/S\equiv (ds/dy)/(dS/dy)$ is of particular interest in the source fireball since both entropy and strangeness yields are nearly preserved in the hadronization process, but the production of strangeness occurs after most of the entropy is created. Up to  a well studied proportionality factor, $s/S$ is the ratio of strange quark abundance to total quark and gluon abundance which is making up the entropy in the bulk. Therefore, $s/S$  measures the degree of chemical equilibration attained in the  {\small QGP}  We observe a constant value of $s/S\simeq 0.03$ (see table~\ref{tab:parameters}), which is in agreement with theoretical expectations for the strange quark mass   $m_s\simeq 100$\,MeV~\cite{Kuznetsova:2006bh}.

\vskip 0.2cm
\noindent\textbf{Comments and Conclusions:} The {\small LHC}2760 experimental environment has opened  a new opportunity to investigate in detail  the  hadron production mechanisms.  Precise particle tracking near to interaction vertex in the {\small ALICE} removes the need for off-line corrections of weak interaction decays, and at the same time, vertex tracking   enhances the efficiency of track identification,  increasing considerably the precision of particle yield measurement~\cite{Abelev:2012vx,QM2012b}.  All  {\small LHC} experimental results used in the present work were obtained in this way  by the {\small ALICE} experiment  for Pb--Pb collisions at $\sqrt{s_{NN}}=2.76$ TeV,  limited to the  central unit of rapidity interval $-0.5<y<0.5$.

In this new experimental environment we show the necessity to introduce the final state hadron chemical non-equilibrium, which describes well all experimental results obtained in the  Pb--Pb collisions at $\sqrt{s_{NN}}=2.76\,\mathrm{TeV}$ from {\small LHC}. As figure~\ref{fig:firstfit} shows,  $\gamma_q\simeq 1.6$ (non-equilibrium of light quarks) allows to describe  the ratio p$/\pi = 0.046\pm0.003$~\cite{Abelev:2012vx,Abelev:2012wca} together with yields of multi-strange baryons $\Xi$ and $\Omega$. 

Another approach to describe the data including the  `anomalous' proton yield at {\small LHC} involves chemical equilibrium hadronization at relatively high $T$ followed by hadron interactions~\cite{Steinheimer:2012rd,Karpenko:2012yf}. We note the chemical equilibrium {\small SHM} yields at hadronization in  \rf{fig:firstfit}, which  overpredict  proton yield, and at the same time underpredict  both $\Xi$ and $\Omega$. Any alternate data explanation   must come to terms with this situation, thus it must deplete protons and enhance both  $\Xi$ and $\Omega$, and at the same time the ratio p$/\pi$ must remain practically constant . This is difficult, as we now discuss, looking closer at the results of Ref.~\cite{Steinheimer:2012rd,Karpenko:2012yf}:\\
$\bullet$   we see in figure 1 of~\cite{Steinheimer:2012rd} that if and when equilibrium style hadronization occurs and leads to high $T$ the necessary post-hadronization reactions  deplete protons and $\Xi$  and enhance  of $\Omega$. This means that the already too small a yield of $\Xi$ is further depleted and disagrees gravely with experiment.\\
$\bullet$   the measured p$/\pi$ ratio in the 0--5\% centrality bin can be made consistent with post-hadronization proton--anti-proton annihilation~\cite{Karpenko:2012yf}. This fine tunes  model parameters and as a result for the 20--30\% centrality bin Ref.~\cite{Karpenko:2012yf} reports increased    p$/\pi=0.058$.\\
$\bullet$   the model predicts for peripheral collisions yet  less annihilation and thus a  p$/\pi$ ratio approaching equilibrium SHM value, which is twice as large as experiment. While  experiment for p$/\pi$  seen in \rf{fig:ratios} is a constant for all centralities,   Ref.~\cite{Karpenko:2012yf} thus predicts a rapid variation by about factor of two. 


These arguments lead to the conclusion that post-hadronization interactions are inconsistent with the experimental data of baryon yields at {\small LHC}. On the other hand, our  chemical non-equilibrium {\small SHM} at {\small LHC} produces a high confidence level fit  $\chi^2/\mathrm{ndf}=7.4/8<1$. Prior {\small SPS} and {\small RHIC} data analysis~\cite{Rafelski:2004dp,Letessier:2005qe,Petran:2011aa} has already  strongly favored chemical non-equilibrium variant of {\small SHM}. The implied sudden hadronization picture is perfectly consistent with the anisotropic flow of quarks leading to the final hadron momentum distribution  azimuthal  asymmetry (see e.g.~\cite{Heinz:2013th}). 

Moreover, we find that LHC and RHIC results are quite consistent in our approach, we obtain the same hadronization condition ($\varepsilon$, $P$, $\sigma$) at {\small LHC} as previously reported at {\small RHIC}, which in turn agrees with high energy {\small SPS}~\cite{Letessier:2005qe}. The energy density of hadronizing matter is $0.50\pm0.05$ GeV/fm$^3$, which is about 3.3 times the energy density of nuclear matter, the pressure is $P=82\pm8\,\text{MeV/fm}^3=(158\pm 4\,\text{MeV})^4$,  as is seen in the bottom part of \rf{fig:entropy}, and which has been proposed in~\cite{Rafelski:2009jr}. The bottom part of table~\ref{tab:parameters} also shows that the entropy density is constant: $\sigma=3.35\pm0.30\,\mathrm{fm}^{-3}$ for both experiments and all centralities.

These typical {\small QGP} properties, including  $s/S\to  0.03$, mean that  at {\small LHC}, the source of hadrons is a chemically equilibrated strangeness saturated {\small QGP} fireball. Furthermore, the universal hadronization condition cannot be viewed anymore as being due to successive particle emission, or to proceed via equilibrated hadron gas phase.

\vskip 0.2cm
\noindent\textbf{Acknowledgments:}  Work  supported by a grant from the U.S. Department of Energy, DE-FG02-04ER41318.


\end{document}